\documentclass[12pt]{article}
\PassOptionsToPackage{noadjust}{cite}
\usepackage{graphicx}
\usepackage{verbatim}
\usepackage{fancybox}
\usepackage{hep}
\usepackage{wrapfig}
\usepackage{textpos}
\usepackage{ifthen} 
\usepackage{hyperref}
\usepackage{hepparticles}
\usepackage{mathrsfs}
\usepackage{amsmath}
\usepackage{amssymb}
\usepackage{amsthm}
\usepackage{mathrsfs}
\usepackage{multirow}
\usepackage{xspace}
\usepackage{maybemath}
\usepackage{color}
\definecolor{gray}          {cmyk}{0   , 0   , 0   , 0.50}
\usepackage{multirow}
\newboolean{uprightparticles}
\setboolean{uprightparticles}{false} 

\newcommand{\pD}{\HepParticle{D}{}{0} }

\newcommand{\pDbar}{\HepAntiParticle{D}{}{0} }

\newcommand{\ppislow}{\HepParticle{\pi}{s}{+} }

\newcommand{\pB}{\HepParticle{B}{}{} }

\newcommand{\deltam}{$\Delta m$ }

\newcommand{\AGamma}{$A_{\Gamma}$}

\def\ux85 {\mbox{UX85}\xspace}

\ifthenelse{\boolean{uprightparticles}}
{

 \def\Ppi         {\ensuremath{\uppi}\xspace}

 \def\PDelta      {\ensuremath{\Delta}\xspace}                 
 \def\PXi      {\ensuremath{\Xi}\xspace}                 
 \def\PLambda      {\ensuremath{\Lambda}\xspace}                 
 \def\PSigma      {\ensuremath{\Sigma}\xspace}                 
 \def\POmega      {\ensuremath{\Omega}\xspace}                 
 \def\PUpsilon      {\ensuremath{\Upsilon}\xspace}                 
 
 \def\PB      {\ensuremath{\mathrm{B}}\xspace}                 
                  
 \def\PD      {\ensuremath{\mathrm{D}}\xspace}

 \def\PK      {\ensuremath{\mathrm{K}}\xspace}

 \def\Pi      {\ensuremath{\mathrm{i}}\xspace}

}
{

 \def\Ppi         {\ensuremath{\pi}\xspace}

 \mathchardef\PDelta="7101
 \mathchardef\PXi="7104
 \mathchardef\PLambda="7103
 \mathchardef\PSigma="7106
 \mathchardef\POmega="710A
 \mathchardef\PUpsilon="7107
                  
 \def\PB      {\ensuremath{B}\xspace}                 
                  
 \def\PD      {\ensuremath{D}\xspace}

 \def\PK      {\ensuremath{K}\xspace}

 \def\Pi      {\ensuremath{i}\xspace}

}

\def\pion  {\ensuremath{\Ppi}\xspace}

\def\pipi  {\ensuremath{\pion^+\pion^-}\xspace}

\def\kaon  {\ensuremath{\PK}\xspace}
  \def\Kbar  {\kern 0.2em\overline{\kern -0.2em \PK}{}\xspace}

\def\Kz    {\ensuremath{\kaon^0}\xspace}
\def\Kzb   {\ensuremath{\Kbar^0}\xspace}
\def\KzKzb {\ensuremath{\Kz \kern -0.16em \Kzb}\xspace}
\def\Kp    {\ensuremath{\kaon^+}\xspace}
\def\Km    {\ensuremath{\kaon^-}\xspace}

\def\KpKm  {\ensuremath{\Kp \kern -0.16em \Km}\xspace}

  \def\Dbar    {\kern 0.2em\overline{\kern -0.2em \PD}{}\xspace}
\def\D       {\ensuremath{\PD}\xspace}

\def\Dz      {\ensuremath{\D^0}\xspace}
\def\Dzb     {\ensuremath{\Dbar^0}\xspace}
\def\DzDzb   {\ensuremath{\Dz {\kern -0.16em \Dzb}}\xspace}
\def\Dp      {\ensuremath{\D^+}\xspace}
\def\Dm      {\ensuremath{\D^-}\xspace}

\def\DpDm    {\ensuremath{\Dp {\kern -0.16em \Dm}}\xspace}

\def\Dstarp  {\ensuremath{\D^{*+}}\xspace}

\def\B       {\ensuremath{\PB}\xspace}
  \def\Bbar    {\kern 0.18em\overline{\kern -0.18em \PB}{}\xspace}

  \def\Y#1S{\ensuremath{\PUpsilon{(#1S)}}\xspace}

\def\Lbar {\ensuremath{\kern 0.1em\overline{\kern -0.1em\PLambda}}\xspace}

\def\to                 {\ensuremath{\rightarrow}\xspace}

\def\AT#1     {\ensuremath{A_{\mathrm{T}}^{#1}}\xspace}           

\def\C#1      {\ensuremath{\mathcal{C}_{#1}}\xspace}                       
\def\Cp#1     {\ensuremath{\mathcal{C}_{#1}^{'}}\xspace}                    
\def\Ceff#1   {\ensuremath{\mathcal{C}_{#1}^{\mathrm{(eff)}}}\xspace}        
\def\Cpeff#1  {\ensuremath{\mathcal{C}_{#1}^{'\mathrm{(eff)}}}\xspace}       
\def\Ope#1    {\ensuremath{\mathcal{O}_{#1}}\xspace}                       
\def\Opep#1   {\ensuremath{\mathcal{O}_{#1}^{'}}\xspace}                    

\def\agamma     {\ensuremath{A_{\Gamma}}\xspace}

\newcommand{\tev}{\ensuremath{\mathrm{\,Te\kern -0.1em V}}\xspace}
\newcommand{\gev}{\ensuremath{\mathrm{\,Ge\kern -0.1em V}}\xspace}
\newcommand{\mev}{\ensuremath{\mathrm{\,Me\kern -0.1em V}}\xspace}
\newcommand{\kev}{\ensuremath{\mathrm{\,ke\kern -0.1em V}}\xspace}
\newcommand{\ev}{\ensuremath{\mathrm{\,e\kern -0.1em V}}\xspace}
\newcommand{\gevc}{\ensuremath{{\mathrm{\,Ge\kern -0.1em V\!/}c}}\xspace}
\newcommand{\mevc}{\ensuremath{{\mathrm{\,Me\kern -0.1em V\!/}c}}\xspace}
\newcommand{\gevcc}{\ensuremath{{\mathrm{\,Ge\kern -0.1em V\!/}c^2}}\xspace}
\newcommand{\gevgevcccc}{\ensuremath{{\mathrm{\,Ge\kern -0.1em V^2\!/}c^4}}\xspace}
\newcommand{\mevcc}{\ensuremath{{\mathrm{\,Me\kern -0.1em V\!/}c^2}}\xspace}

\newcommand{\chisq}{\ensuremath{\chi^2}\xspace}

\def\gsim{{~\raise.15em\hbox{$>$}\kern-.85em
          \lower.35em\hbox{$\sim$}~}\xspace}
\def\lsim{{~\raise.15em\hbox{$<$}\kern-.85em
          \lower.35em\hbox{$\sim$}~}\xspace}

\def\tell1  {TELL1\xspace}
\def\ukl1   {UKL1\xspace}

\def\deltam     {\ensuremath{\Delta m}\xspace}

\def\DDbar   {\ensuremath{\kern -0.1em \stackrel{\kern 0.1em \textsf{\fontsize{5pt}{1em}\selectfont(---)}}{D}\kern -0.3em}\xspace}
\def\AAbar   {\ensuremath{\kern -0.2em \stackrel{\kern 0.2em \textsf{\fontsize{5pt}{1em}\selectfont(---)}}{A}\kern -0.3em}\xspace}
\def\AfAfbar   {\ensuremath{\kern -0.2em \stackrel{\kern 0.2em \textsf{\fontsize{5pt}{1em}\selectfont(---)}}{A}_{\kern -0.3em f}\kern -0.3em}\xspace}

\def\lnIPChisq {\ensuremath{\ln(\mathrm{IP}\chisq)}\xspace}

\def\pbnr{}
\def\speaker{Mark Smith}
\def\onbehalfof{the LHCb collaboration}
\def\title{Measurement of $A_{\Gamma}$}
\def\affiliation{School of Physics and Astronomy\\
The University of Manchester, Manchester, UK}
\def\support{The workshop was supported by the University of Manchester, IPPP, STFC, and IOP}

\newcommand\pubnumber{\pbnr}
\newcommand\pubdate{\today}
\def\Title#1{\begin{center} {\Large #1 } \end{center}}
\def\Author#1{\begin{center}{ \sc #1} \end{center}}

\newcommand{\OnBehalf}[1]{\sbox0{#1}\ifdim\wd0=0pt
        {}
	\else
	{\\on behalf of #1}
	\fi}
\newcommand{\SupportedBy}[1]{\sbox0{#1}\ifdim\wd0=0pt
        {}
	\else
	{\footnote{#1}}
	\fi}
\def\Address#1{\begin{center}{ \it #1} \end{center}}

\newcommand\pubblock{\includegraphics[width=5cm]{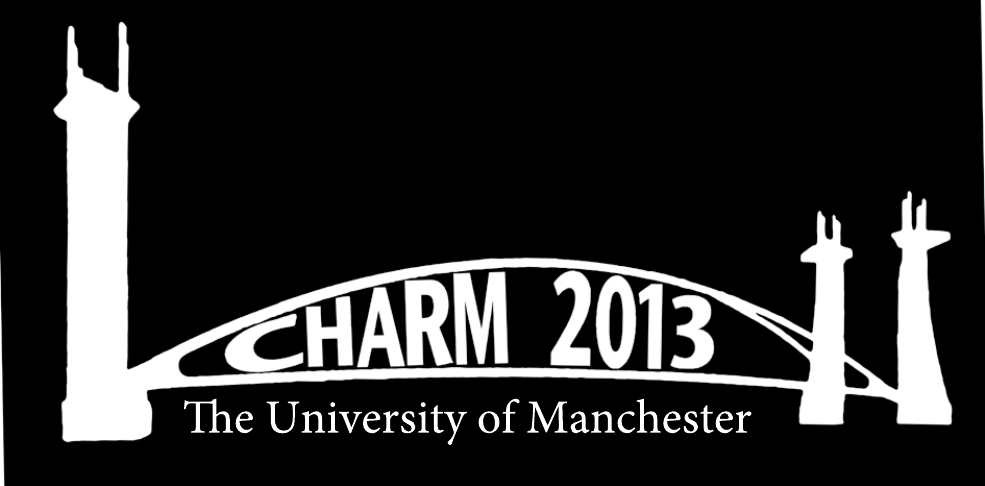}\hfill{\begin{tabular}{l} \pubnumber\\
         \pubdate  \end{tabular}}}
\newenvironment{Abstract}{\begin{quotation}  }{\end{quotation}}
\newenvironment{Presented}{\begin{quotation} \begin{center} 
             PRESENTED AT\end{center}\bigskip 
      \begin{center}\begin{large}}{\end{large}\end{center} \end{quotation}}

\def\venue{The 6$^{th}$ International Workshop on Charm Physics\\
(CHARM 2013)\\
Manchester, UK,  31 August -- 4 September, 2013}

\def\beq{\begin{equation}}
\def\eeq#1{\label{#1}\end{equation}}
\def\eeqn{\end{equation}}

\def\beqa{\begin{eqnarray}}
\def\eeqa#1{\label{#1}\end{eqnarray}}
\def\eeqan{\end{eqnarray}}

\let\bar=\overbar

\def\Dslash{\not{\hbox{\kern-4pt $D$}}}
\def\dslash{\not{\hbox{\kern-2pt $\del$}}}

\def\msb{{\bar{\ssstyle M \kern -1pt S}}}

\begin{document}
\begin{titlepage}
\pubblock
\vfill
\Title{\title}
\vfill
\Author{\speaker\SupportedBy{\support}\OnBehalf{\onbehalfof}}
\Address{\affiliation}
\vfill
\begin{Abstract}
The measurement of the charm CP violation observable $A_{\Gamma}$ using \mbox{1 fb$^{-1}$} of $pp$ collisions at $\sqrt{s}=7$ TeV recorded by the LHCb detector in 2011 is presented. This new result is the most accurate to date.
\end{Abstract}
\vfill
\begin{Presented}
\venue
\end{Presented}
\vfill
\end{titlepage}
\def\thefootnote{\fnsymbol{footnote}}
\setcounter{footnote}{0}
%

\section{Introduction}

CP violation in charm meson decays is expected to be small in the Standard Model (SM) and any significant enhancement would be a signal of New Physics (NP). Thus far no CP violation has been unambiguously observed in the charm system.

The CP violation observable $A_{\Gamma}$ is defined as the asymmetry of the effective lifetimes of \Dz and \Dzb decaying to the same CP eigenstate, \Kp{}\Km or \pipi,
\begin{equation}
A_{\Gamma} = \frac{\hat{\Gamma}(\Dz\to\Kp\Km) - \hat{\Gamma}(\Dzb\to\Kp\Km)}{\hat{\Gamma}(\Dz\to\Kp\Km) + \hat{\Gamma}(\Dzb\to\Kp\Km)} \approx \frac{A_{m}+A_{d}}{2}y\cos\phi-x\sin\phi,
\end{equation}
where $A_{m}$ and $A_{d}$ are the asymmetries due to CP violation in mixing and decay respectively, $\phi$ is the interference phase between mixing and decay and $x$ and $y$ are the charm mixing parameters.

In the Standard Model \AGamma{} is expected to be small\cite{Lenz}($\sim$10$^{-4}$) and roughly independent of the final state. New Physics (NP) models may introduce larger CP violation and some final state dependence of the phase $\phi$ leading to a difference in \AGamma{} between the \Kp{}\Km and \pipi final states\cite{Sokoloff},
\begin{equation}
\Delta A_{\Gamma} = A_{\Gamma}(KK) - A_{\Gamma}(\pi\pi) = \Delta A_{D} y\cos\phi + (A_{M}+A_{D})y\Delta\cos\phi - x\Delta\sin\phi.
\end{equation}

\begin{figure}[htb]
\centering
\includegraphics[width=0.48\textwidth]{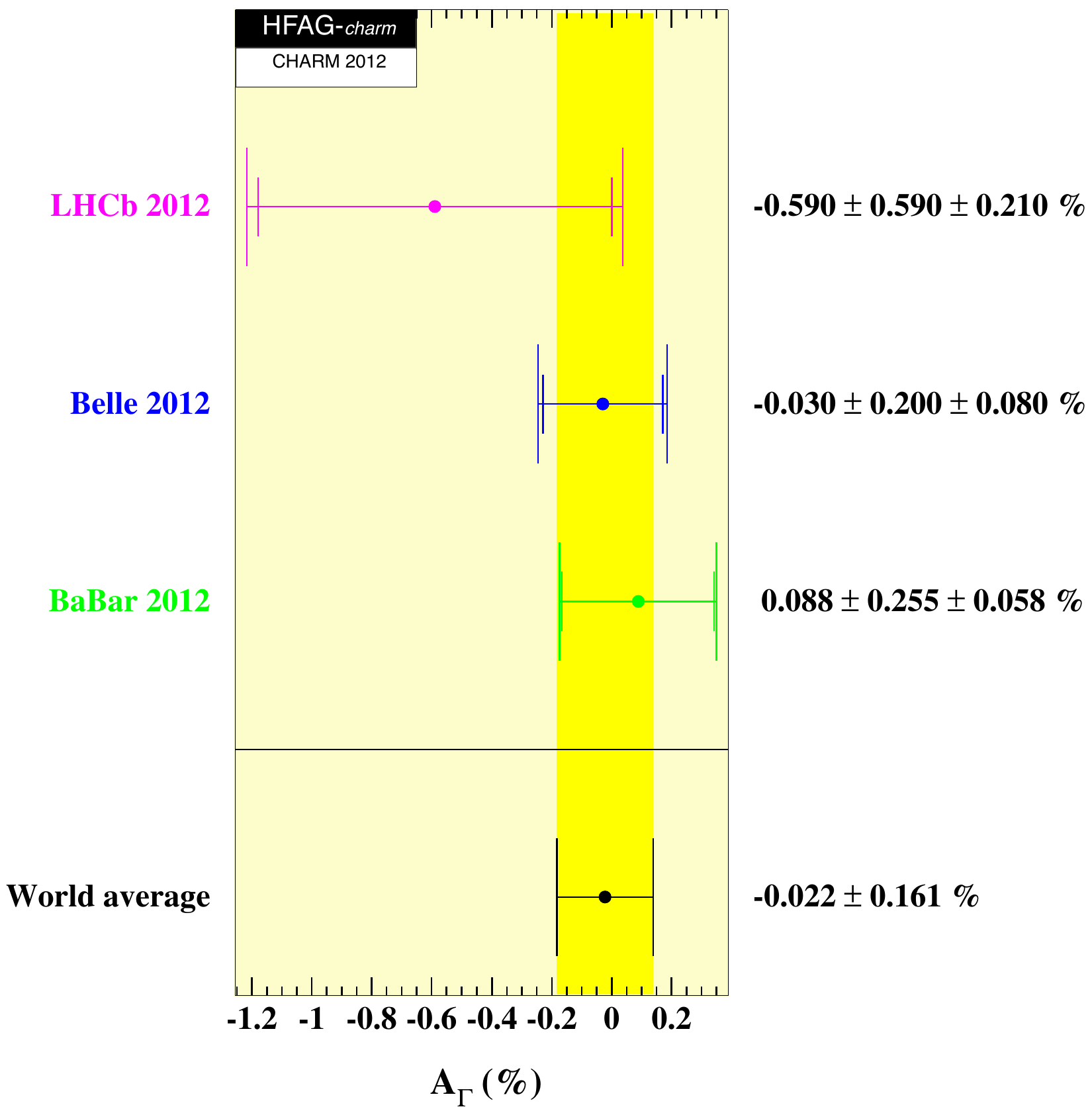}
\caption{Experimental status of \AGamma.}
\label{fig:agamma}
\end{figure}

The experimental status of the measurement of \AGamma, including the Heavy Flavour Averaging Group (HFAG)\cite{HFAG} average and excluding the results presented here, is shown in Fig.~\ref{fig:agamma}.

Presented here are new results for the measurement of \agamma using 1 fb$^{-1}$ of $pp$ collisions at a centre of mass energy of 7 TeV recorded by the LHCb detector in 2011\cite{paper}.

\section{Analysis Method}

The mean lifetimes of the \Dz and \Dzb are extracted via a fit to their decay times.  The data to be fitted is broken into eight subsets. The splits are motivated by the two detector magnet polarities with which data was taken and two separate data-taking periods to account for know differences in detector alignment and calibration . Finally the \Dz and \Dzb candidates have been fitted separately.

The initial flavour of the \Dz is determined by searching for the decay $\Dstarp\to\Dz\ppislow$ where the charge on the pion indicates the flavour. Due to the small $Q$ value of this decay the pion is referred to as slow.

\begin{figure}[htb]
\centering
\begin{tabular}{c c}
\includegraphics[width=0.48\textwidth]{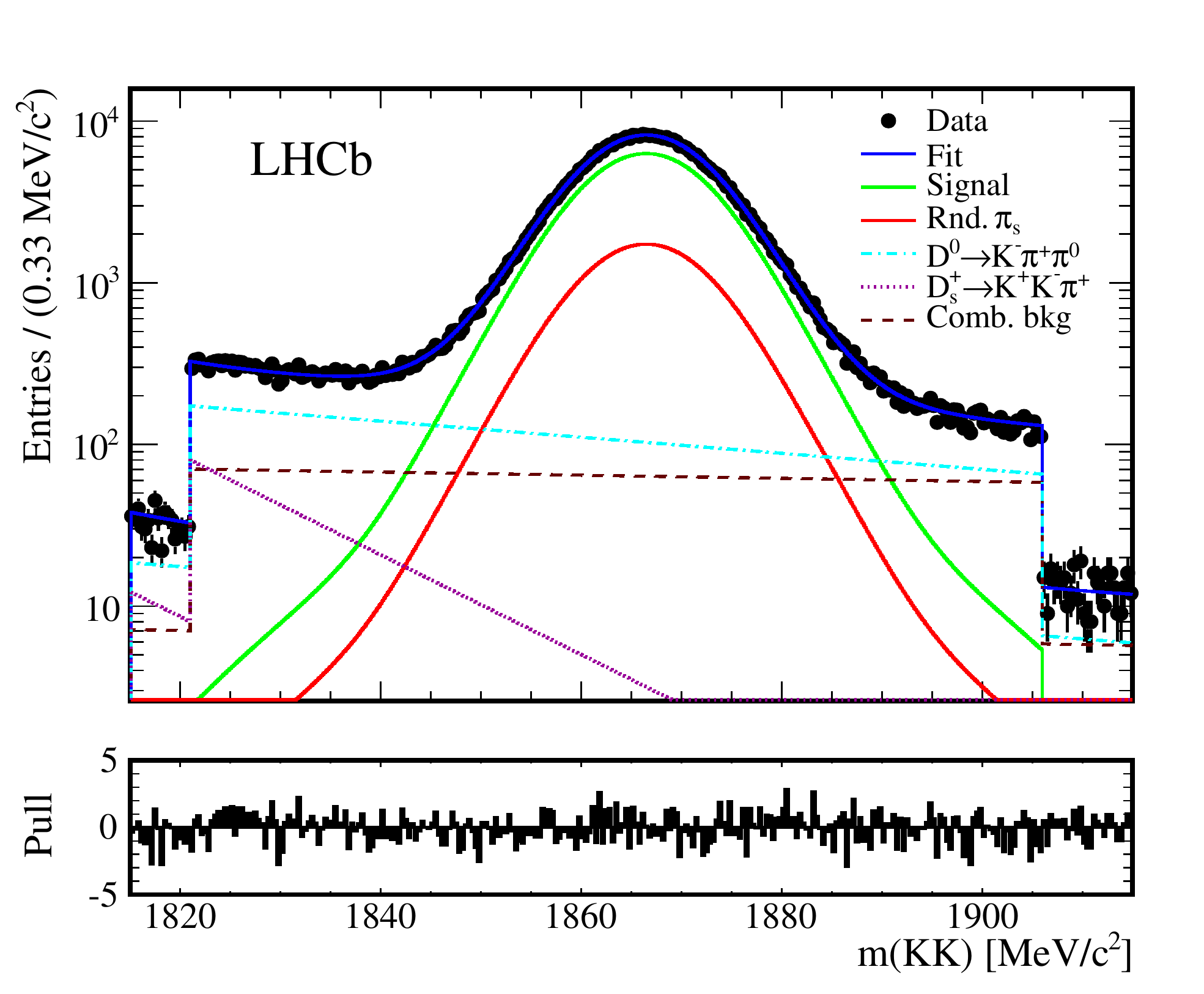}
&
\includegraphics[width=0.48\textwidth]{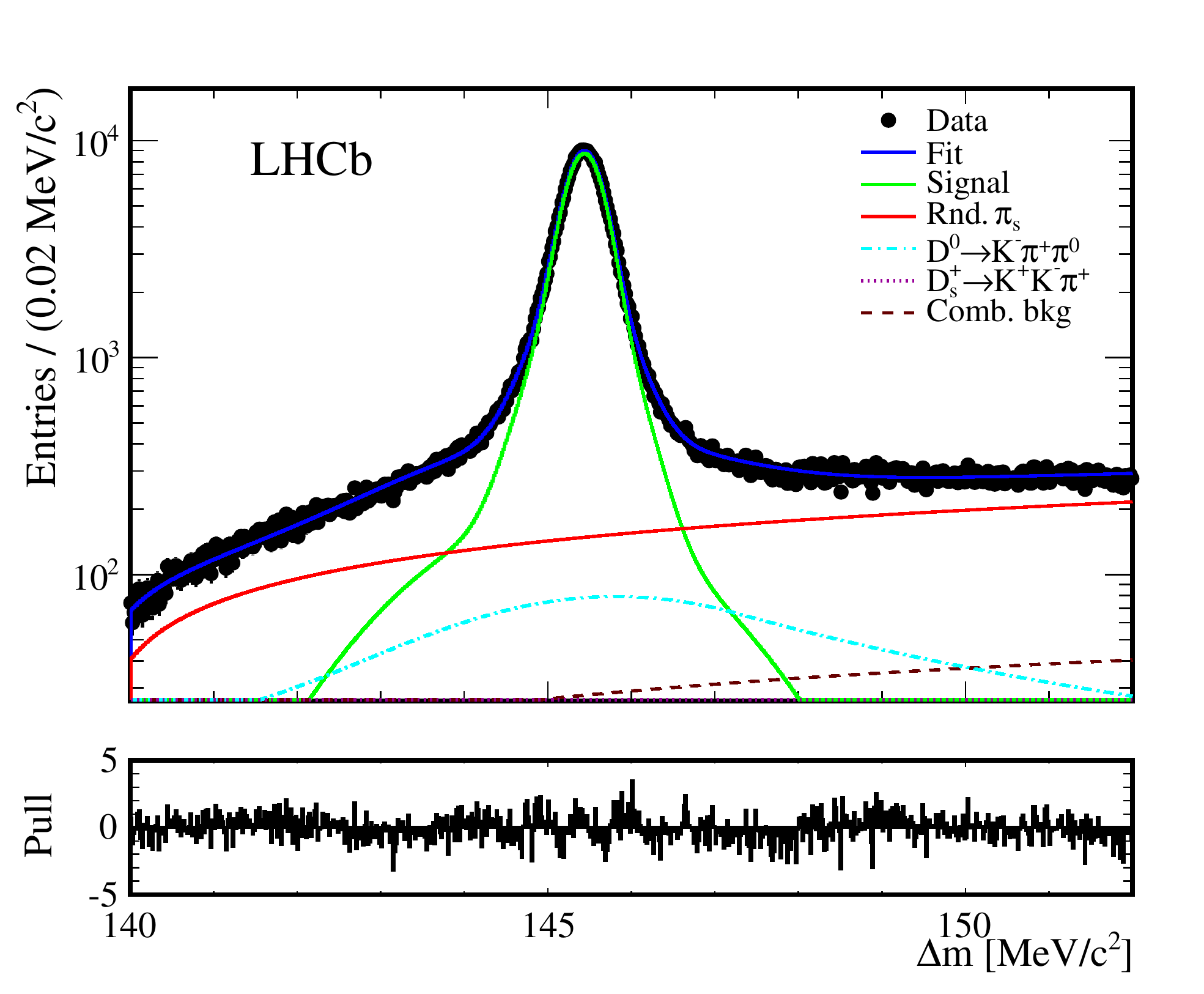}
\end{tabular}
\caption{Fit of the \Dz mass (left) and \deltam (right) for subset of data containing $\Dzb\to\Kp{}\Km$ candidates with magnet polarity down for the earlier run period.}
\label{fig:massfit}
\end{figure}

The procedure is carried out in two stages. In the first the \Dz mass and the difference between the \Dstarp and \Dz masses (\deltam) are fitted simultaneously. This allows for the separation of the signal and background components and the determination of the background probability density functions in the subsequent fits. Example mass and \deltam fit results for the \Kp{}\Km final state can be see in Fig.~\ref{fig:massfit}.

\begin{figure}[htb]
\centering
\begin{tabular}{c c}
\includegraphics[width=0.3\textwidth]{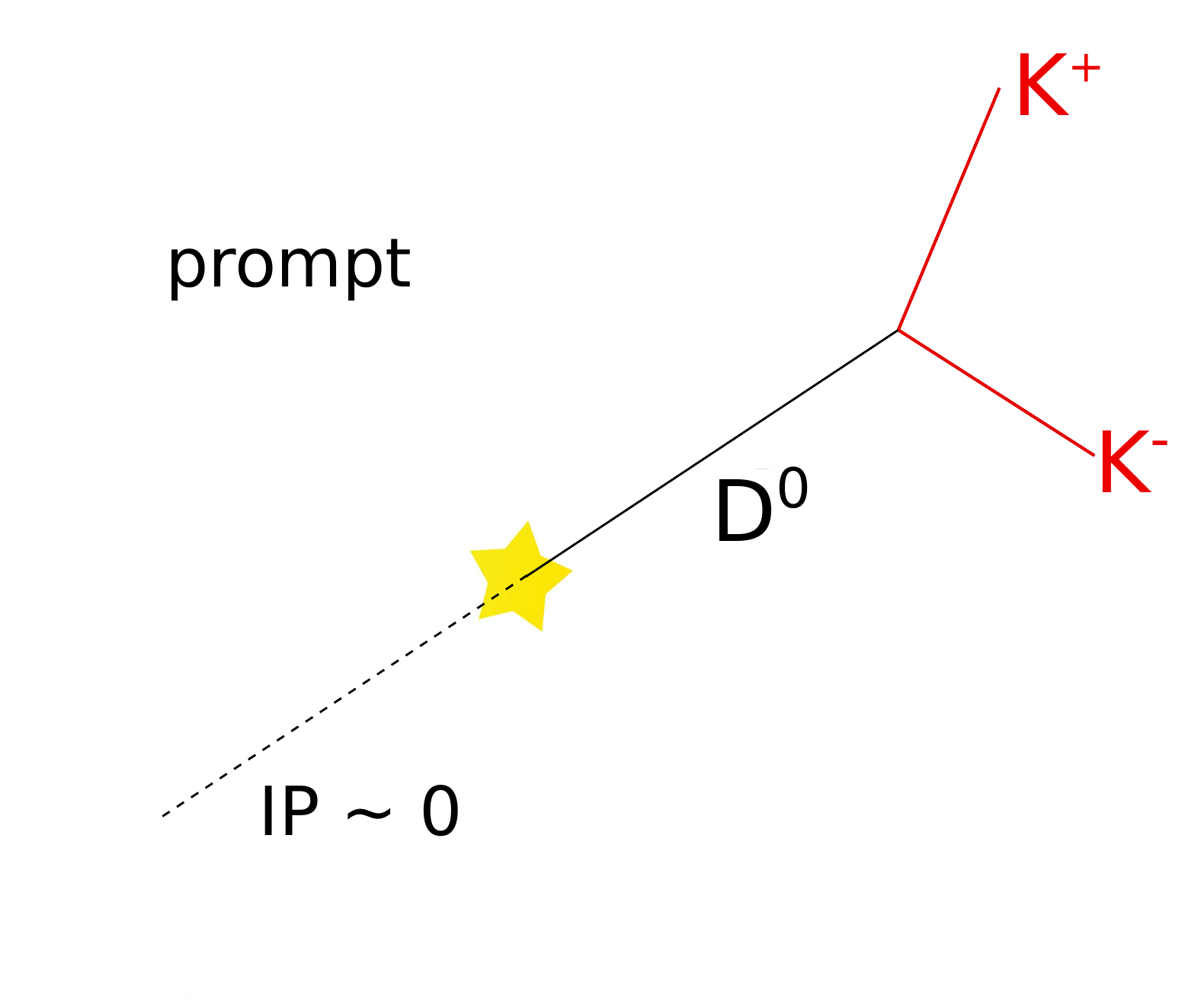}
&
\includegraphics[width=0.3\textwidth]{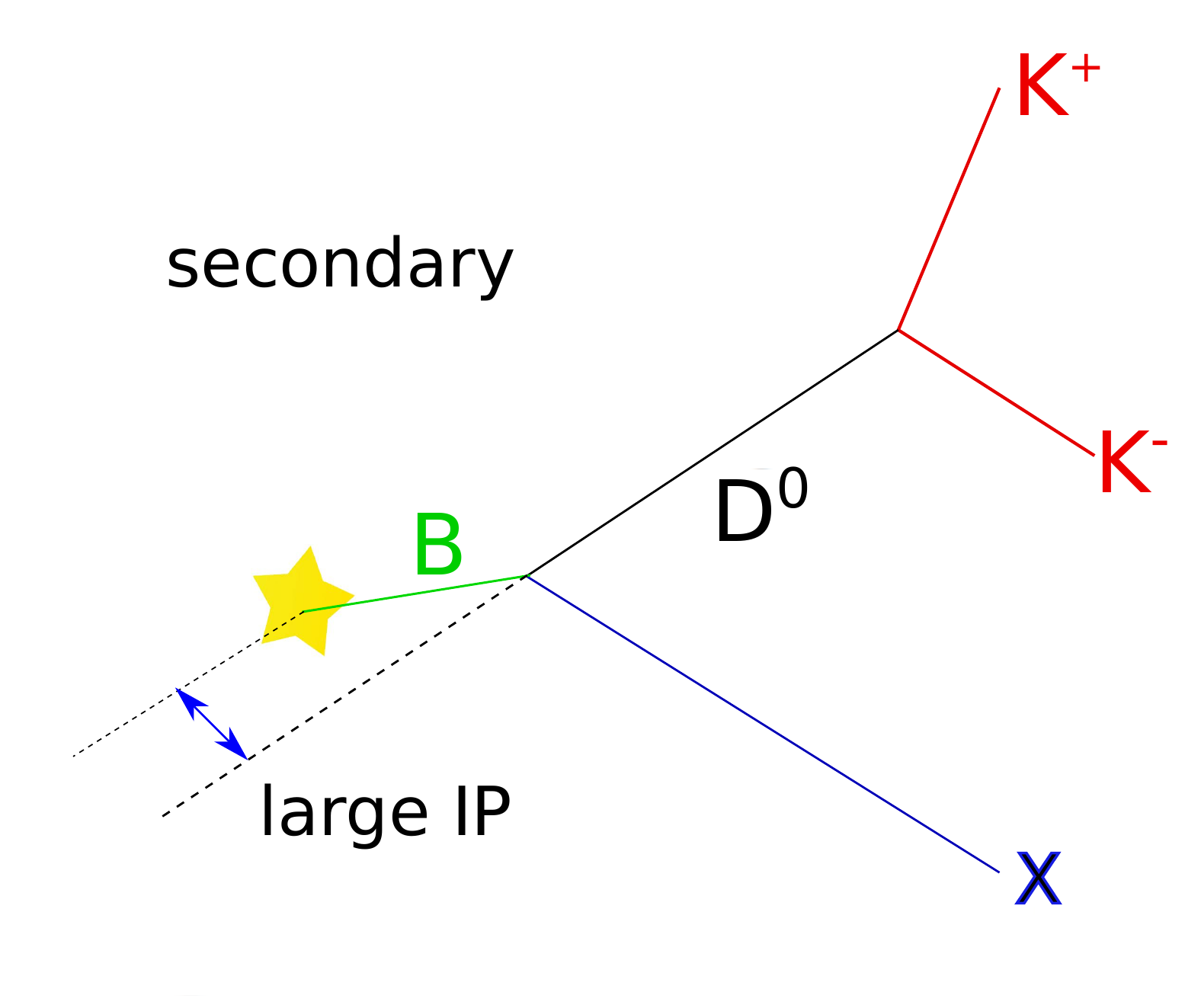}
\end{tabular}
\caption{The separation of prompt (left) and secondary (right) decays by considering their impact parameters.}
\label{fig:secondary}
\end{figure}

\begin{figure}[htb]
\centering
\begin{tabular}{c c}
\includegraphics[width=0.48\textwidth]{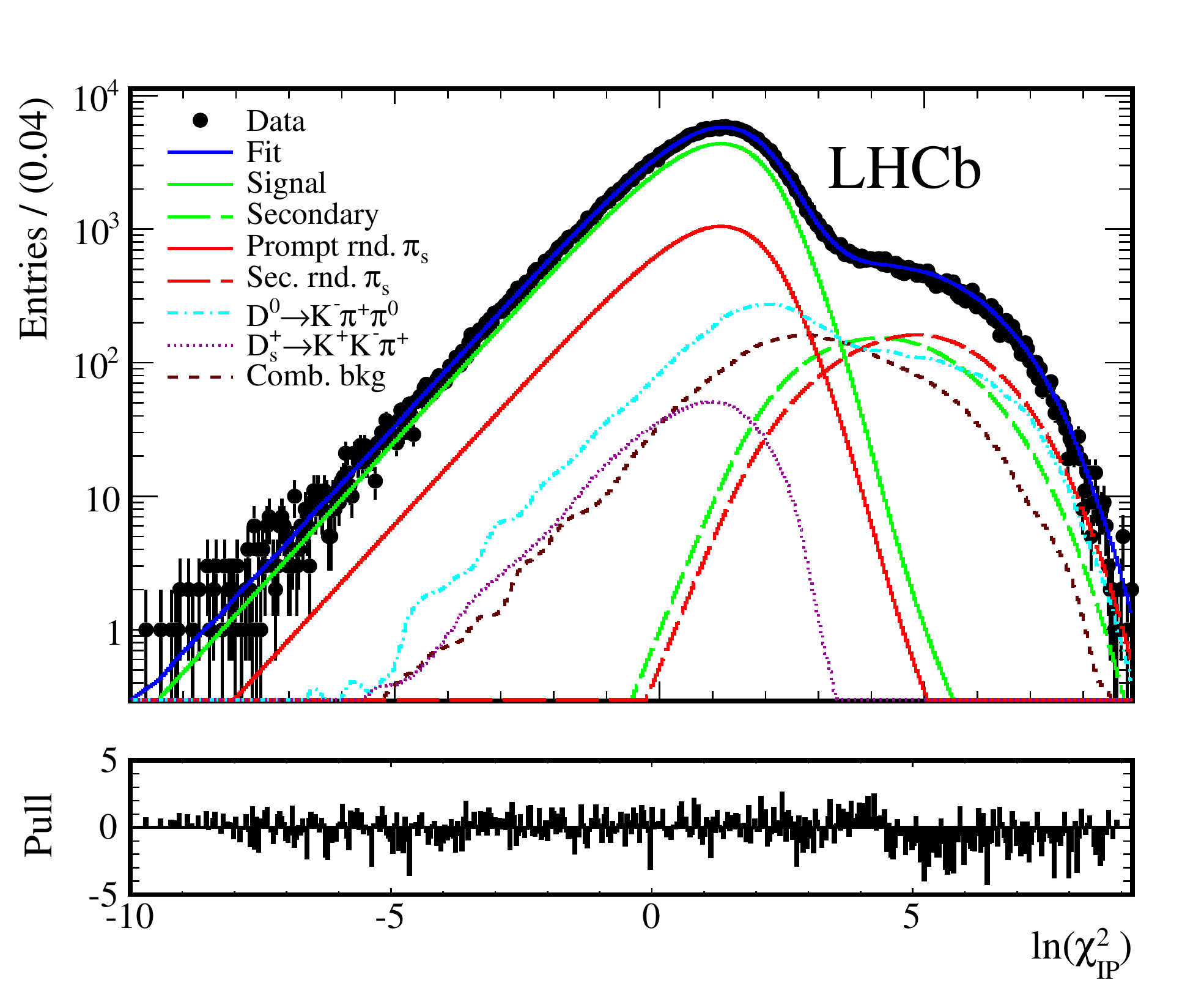}
&
\includegraphics[width=0.48\textwidth]{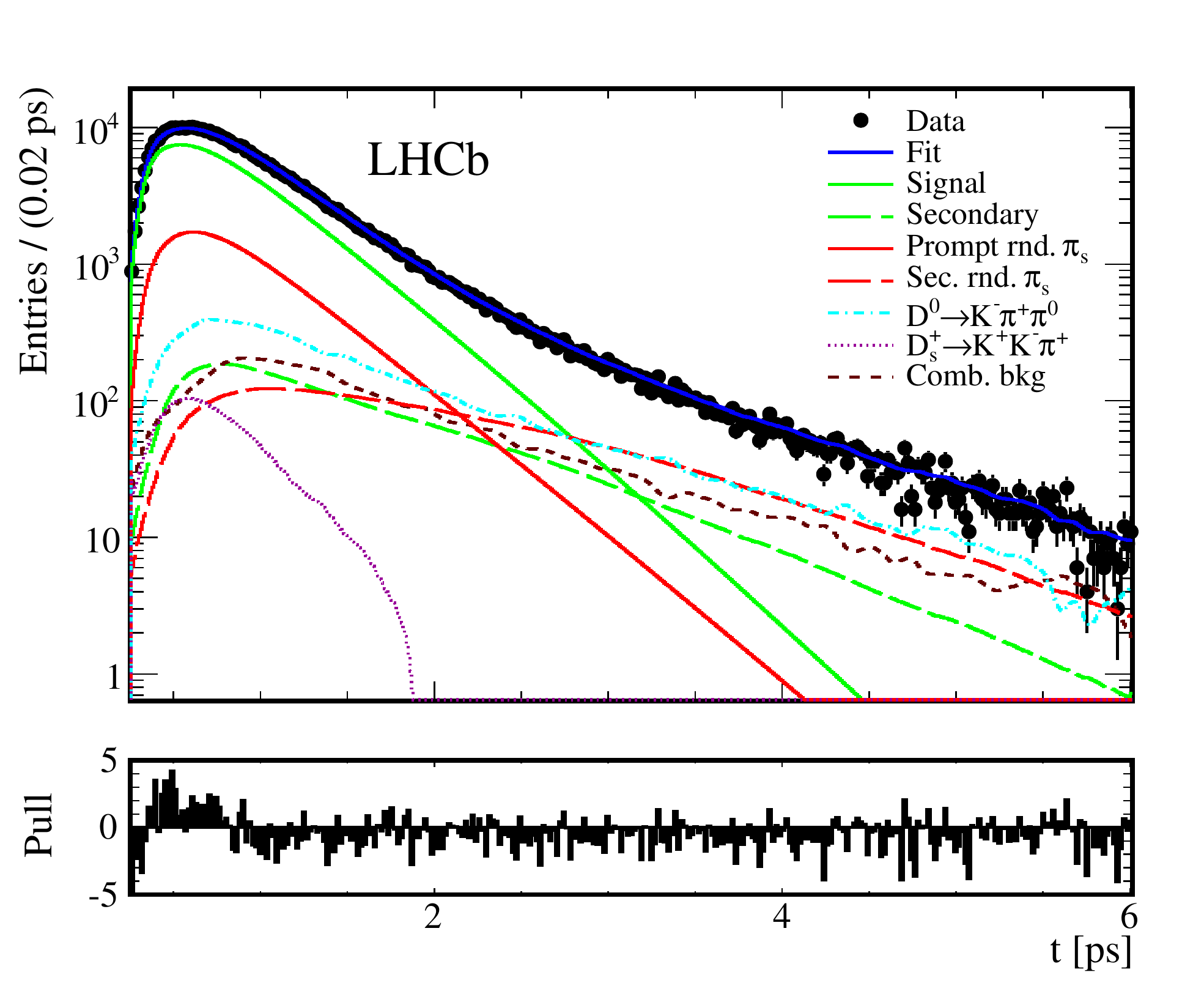}
\end{tabular}
\caption{Fit of the \lnIPChisq (left) and decay time (right) for data subset containing the $\Dzb\to\Kp{}\Km$ candidates with magnet polarity down for the earlier run period.}
\label{fig:timefit}
\end{figure}

The second stage fits \Dz decay times and the natural logarithm of the \Dz impact parameter $\chi^{2}$ (\lnIPChisq). Those \Dz candidates originating from \B decays (secondary) have longer measured lifetimes than those originating at the primary vertex (prompt) as the \B has not been reconstructed. It is therefore necessary to separate these in the fit to avoid biasing the lifetime measurement. This is done using the \lnIPChisq variable. Due to the flight distance of the \B the impact parameter of the \Dz is larger than those of prompt candidates as shown in Fig.~\ref{fig:secondary}. Example fits are in Fig.~\ref{fig:timefit}.

\begin{figure}[htb]
\centering
\begin{tabular}{c c c}
\includegraphics[width=0.3\textwidth]{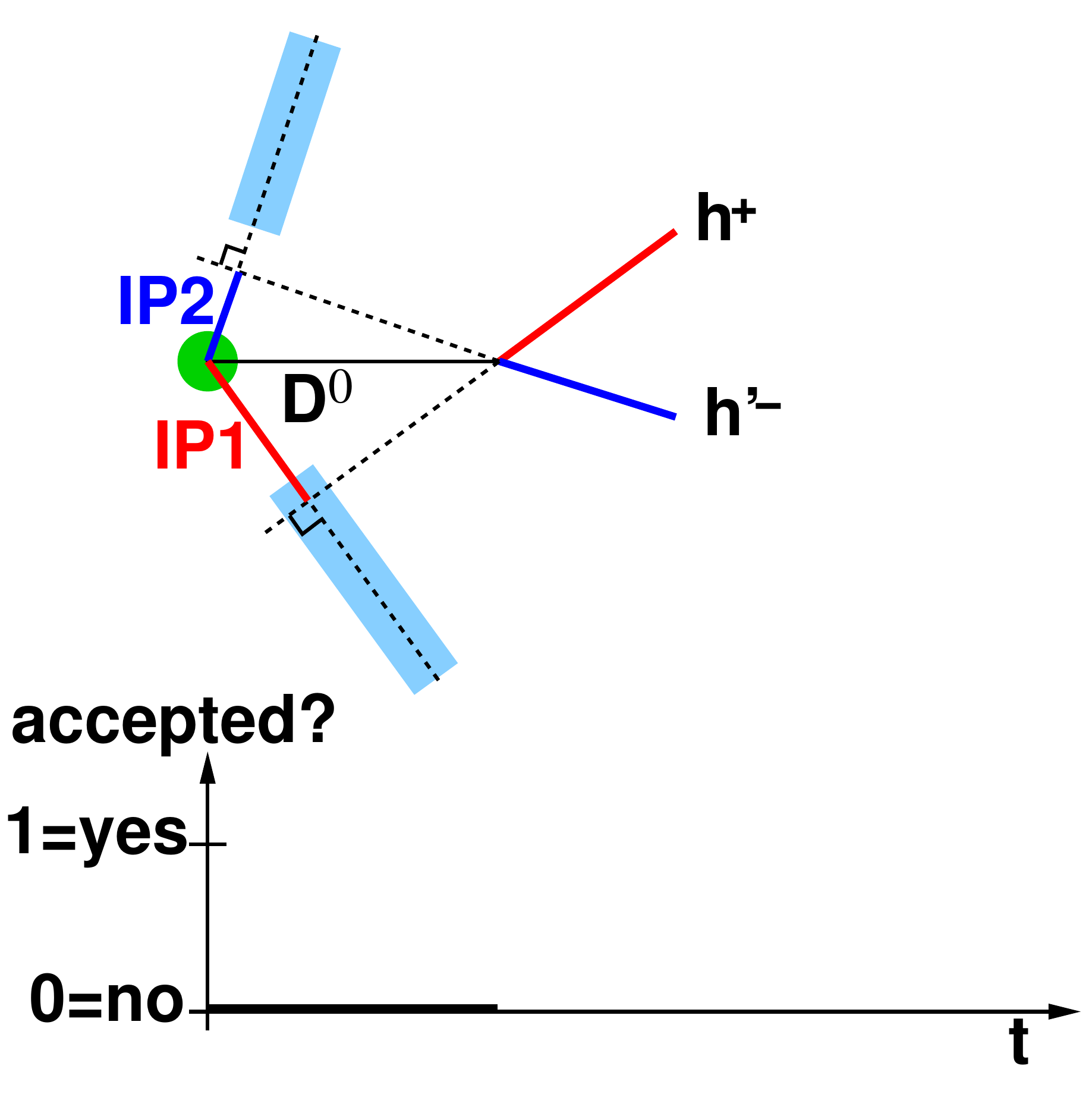}
&
\includegraphics[width=0.3\textwidth]{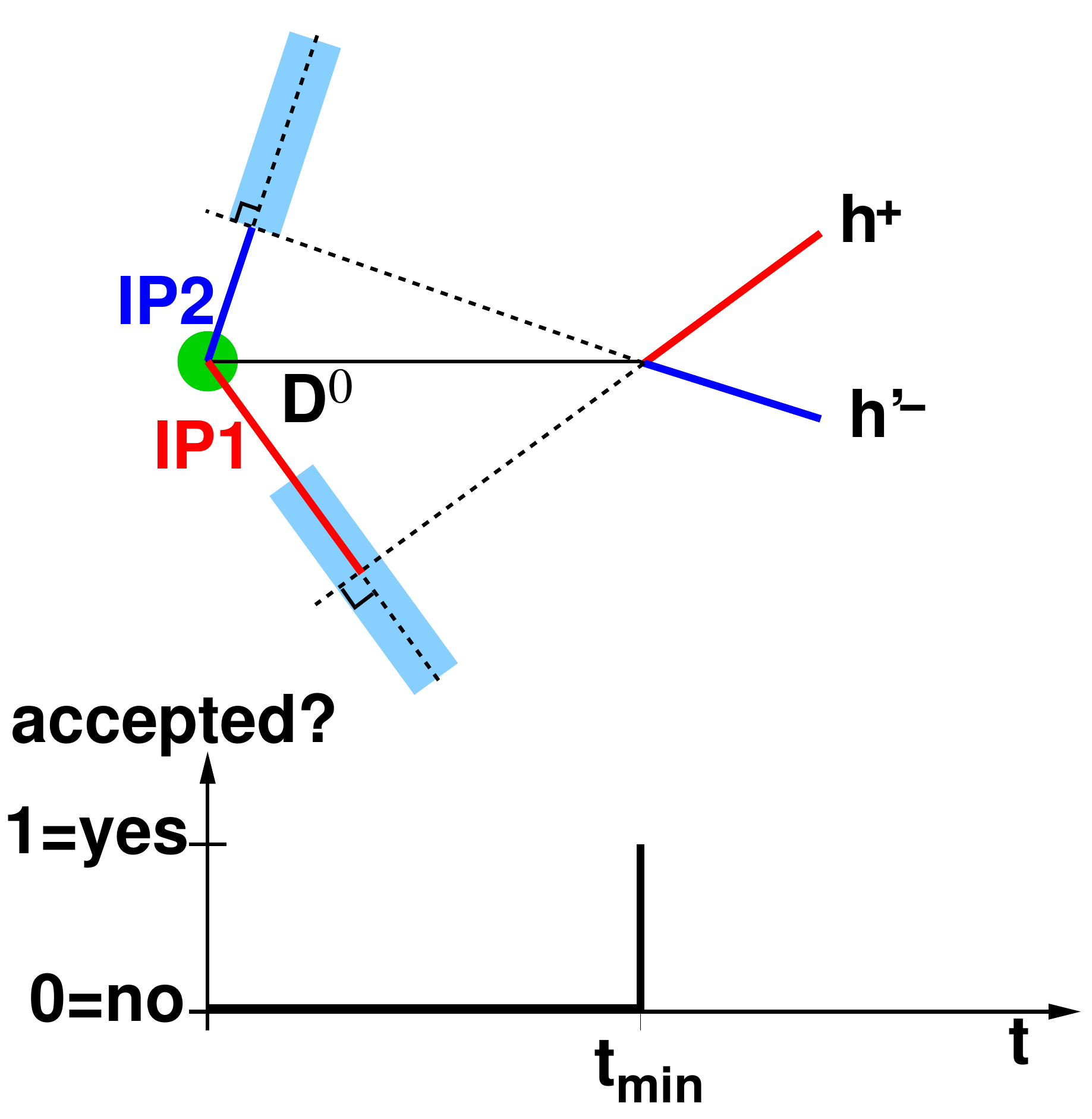}
&
\includegraphics[width=0.3\textwidth]{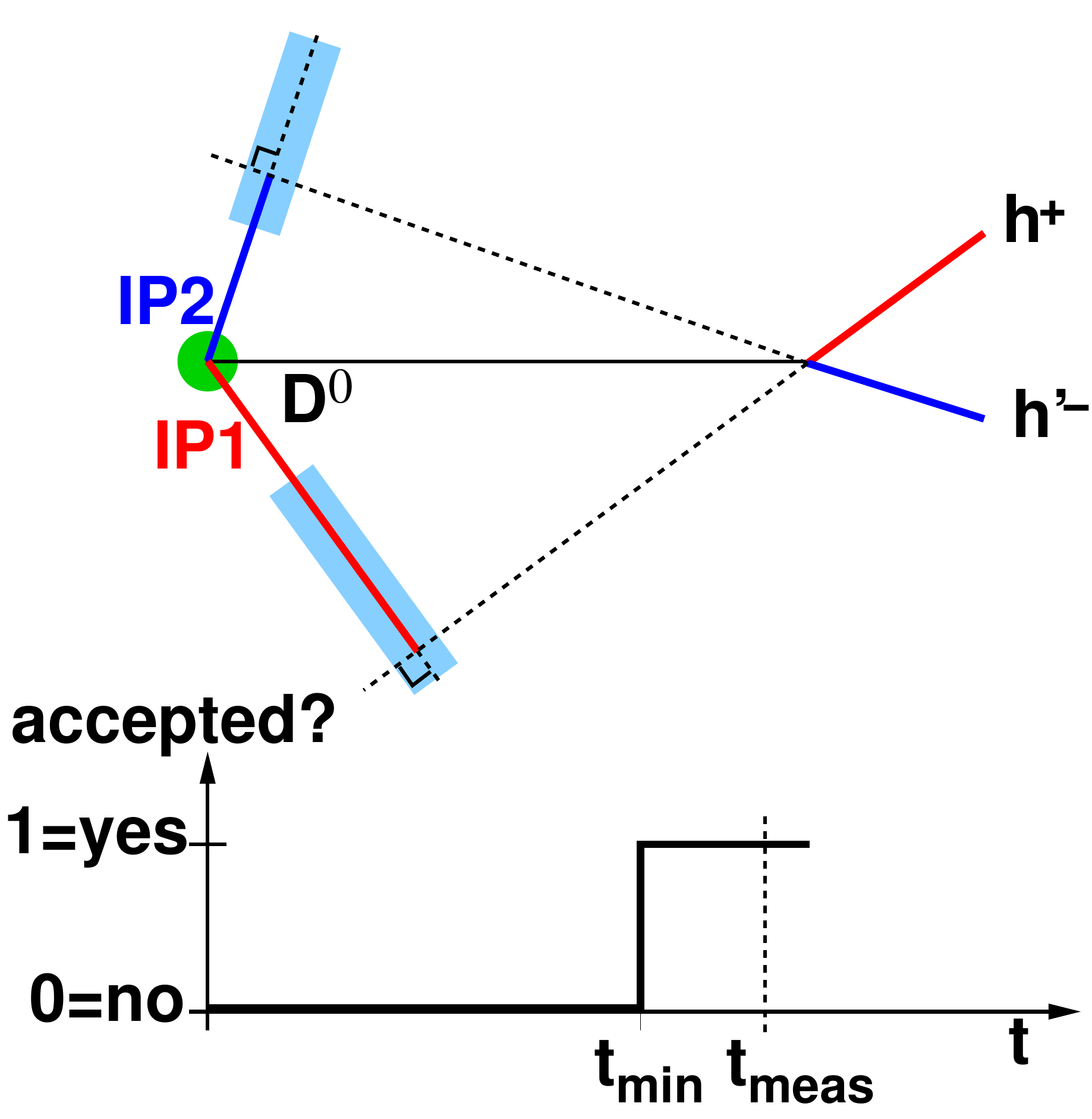}
\end{tabular}
\caption{The swimming method. The \Dz primary vertex is `swum' along the \Dz direction (from left to right). The trigger is rerun for each position and the lifetime of the candidate at which it becomes accepted by the trigger is found (middle).}
\label{fig:swimming}
\end{figure}

Lifetime biases due to the acceptance of the trigger and selections are corrected for using the ``swimming'' method. The \Dz primary vertex is moved along the direction of its flight and the trigger rerun to find the point in \Dz lifetime at which the candidate changes from being rejected to accepted. One can thus construct an acceptance function in \Dz lifetime for each event as shown in Fig.~\ref{fig:swimming}. An average acceptance function for the whole data set can then be constructed and folded in to the fit. For a complete description see \cite{Vava}.

\section{Summary of systematic uncertainties}

The systematic uncertainties of the method are evaluated through a mixture of studies on simplified simulated data and variations to the fit. Table~\ref{tab:systematics} summarises the results of these studies. Additionally some extra considerations such as detector resolution and track reconstruction efficiency (amongst others) are looked at and found to have a negligible effect on the resultant \agamma measurement. The data are also split into bins of various kinematic variables (for example \Dz momentum $p$, transverse momentum $p_{T}$ and flight direction) and no systematic variation in the result is found.

The dominant systematic uncertainty comes from the acceptance function. This includes the uncertainty of the turning point positions determined by the swimming method and their subsequent utilisation in the fit procedure.
\begin{table}[t]
\begin{center}
\begin{tabular}{c|cc}  
Effect & \agamma(\Kp{}\Km)$\times 10^{-3}$ & \agamma(\pipi)$\times 10^{-3}$\\
\hline
Mis-reconstructed bkg. & $\pm 0.02$ & $\pm 0.00$\\
Charm from \pB & $\pm 0.07$ & $\pm0.07$\\
Other backgrounds & $\pm0.02$ & $\pm0.04$\\
Acceptance function & $\pm0.09$ & $\pm0.11$\\
\hline
Total & $\pm0.12$ & $\pm0.14$\\

\end{tabular}
\caption{Summary of the systematic uncertainties on the measurement of \agamma for the two final states.}
\label{tab:systematics}
\end{center}
\end{table}

\section{Results}
The results of the \agamma measurement for the \Kp{}\Km and \pipi final states are:
\begin{equation}
A_{\Gamma}(KK) = (0.35\pm0.62_{stat}\pm0.12_{syst})\times 10^{-3}
\end{equation}
\begin{equation}
A_{\Gamma}(\pi\pi) = (0.33\pm1.06_{stat}\pm0.14_{syst})\times 10^{-3}
\end{equation}

The two numbers show no CP violation within the experimental uncertainty and are consistent with each other. They show a considerable improvement in accuracy over previous results. At the same time a complimentary measurement of \agamma was made on the same data using an alternative method by which the time evolution of the ratio of \pD and \pDbar yields was examined. The two methods yielded consistent results. Analysis of the 2 fb$^{-1}$ 2012 data set is to follow which will increase the precision further.

\section{Acknowledgements}

\noindent We express our gratitude to our colleagues in the CERN
accelerator departments for the excellent performance of the LHC. We
thank the technical and administrative staff at the LHCb
institutes. We acknowledge support from CERN and from the national
agencies: CAPES, CNPq, FAPERJ and FINEP (Brazil); NSFC (China);
CNRS/IN2P3 and Region Auvergne (France); BMBF, DFG, HGF and MPG
(Germany); SFI (Ireland); INFN (Italy); FOM and NWO (The Netherlands);
SCSR (Poland); MEN/IFA (Romania); MinES, Rosatom, RFBR and NRC
``Kurchatov Institute'' (Russia); MinECo, XuntaGal and GENCAT (Spain);
SNSF and SER (Switzerland); NAS Ukraine (Ukraine); STFC (United
Kingdom); NSF (USA). We also acknowledge the support received from the
ERC under FP7. The Tier1 computing centres are supported by IN2P3
(France), KIT and BMBF (Germany), INFN (Italy), NWO and SURF (The
Netherlands), PIC (Spain), GridPP (United Kingdom). We are thankful
for the computing resources put at our disposal by
Yandex LLC (Russia), as well as to the communities behind the multiple open
source software packages that we depend on.


\begin{thebibliography}{99}



\bibitem{Lenz}
M. Bobrowski, A. Lenz, J. Riedl, and J. Rohrwild, \textit{How large can the SM contribution to CP violation in $\Dz-\Dzb$ mixing be?}, JHEP \textbf{03} (2010) 009.

\bibitem{Sokoloff}
A. L. Kagan and M. D. Sokoloff, \textit{On indirect CP violation and implications for $D^0-\bar{D}^0$ and $B^0_s-\bar{B}^0_s$ mixing}, Phys. Rev. \textbf{D80} (2009) 076008.

\bibitem{HFAG}
Heavy Flavor Averaging Group, Y. Amhis \textit{et al.}, \textit{Averages of b-hadron, c-hadron, and $\tau$-lepton
properties as of early 2012}, \textbf{arXiv:1207.1158},updated results and plots available at
\textbf{http://www.slac.stanford.edu/xorg/hfag/}.

\bibitem{paper}
LHCb collaboration, R. Aaij \textit{et al.}, \textit{Measurements of indirect CP asymmetries in $D^0\to
                        K^-K^+$ and $D^0\to\pi^-\pi^+$ decays}, submitted to Phys. Rev. Lett., \textbf{arXiv:1310.7201}.


\bibitem{Vava}
V. Gligorov \textit{et al.}, \textit{Swimming: A data driven acceptance correction algorithm}, J. Phys. Conf. Ser. \textbf{396} (2012) 022016.


\end{thebibliography}
\end{document}